\documentclass[aps,bibnotes,twocolumn]{revtex4}
\usepackage[tbtags]{amsmath}
\usepackage{amssymb}

\newcommand{\beq}{\begin{equation}}
\newcommand{\eeq}{\end{equation}}
\newcommand{\eeql}[1]{\label{#1}\end{equation}}

\begin{document}

\title{Comment on `Aharonov--Casher-Effect Suppression\\ of Macroscopic Tunneling of Magnetic Flux'}

\author{Alec \surname{Maassen van den Brink}}
\email{alec@dwavesys.com}
\affiliation{D-Wave Systems Inc., 320-1985 West Broadway, Vancouver, BC, V6J 4Y3, Canada}

\date{\today}

\maketitle

A recent Letter~\cite{FA} has proposed a device, consisting of an rf SQUID with the junction replaced by a double one, i.e.\ a Bloch transistor. For symmetric Josephson couplings $E_1=E_2$, and for specific flux bias $\Phi_\mathrm{x}$ and gate voltage $V_\mathrm{g}$ applied to the transistor's island, the tunneling of encircled flux $\Phi$ between its two local potential minima is claimed to be suppressed completely due to destructive interference. This comment purports to show that this claim in general is not valid. Namely, the very Hamiltonian~(1) in~\cite{FA}, from which the effect is derived, describes the circuit only under an additional condition.

\setlength{\unitlength}{1mm}
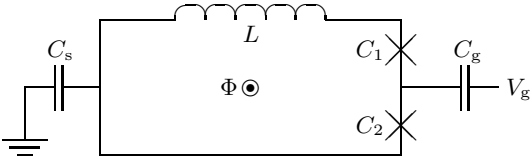
\begin{figure}[h]
\begin{picture}(70,19)
  \multiput(25,18)(4,0){5}{\oval(4,4)[t]}
  \put(32,15){$L$}
  \put(43,18){\line(1,0){10}}
  \put(53,0){\line(0,1){18}}
  \put(51,16){\line(1,-1){4}}
  \put(51,12){\line(1,1){4}}
  \put(47,13){$C_1$}
  \put(51,6){\line(1,-1){4}}
  \put(51,2){\line(1,1){4}}
  \put(47,3){$C_2$}
  \put(53,9){\line(1,0){8}}
  \put(61,6){\line(0,1){6}}
  \put(62,6){\line(0,1){6}}
  \put(60,13){$C_\mathrm{g}$}
  \put(62,9){\line(1,0){4}}
  \put(67,8){$V_\mathrm{g}$}
  \put(23,18){\line(-1,0){10}}
  \put(13,0){\line(0,1){18}}
  \put(13,0){\line(1,0){40}}
  \put(33,9){\circle*{1}}
  \put(33,9){\circle{2}}
  \put(29,8){$\Phi$}
  \put(13,9){\line(-1,0){5}}
  \put(8,6){\line(0,1){6}}
  \put(7,6){\line(0,1){6}}
  \put(6,13){$C_\mathrm{s}$}
  \put(7,9){\line(-1,0){4}}
  \put(3,9){\line(0,-1){7}}
  \put(0,2){\line(1,0){6}}
  \put(1,1){\line(1,0){4}}
  \put(2,0){\line(1,0){2}}
\end{picture}
\caption{A superconducting ring including a Bloch transistor.}
\label{fig1}
\end{figure}

For definiteness, we model the circuit as in Fig.~\ref{fig1}. The stray capacitance $C_\mathrm{s}$ is formally necessary if $V_\mathrm{g}$ is to have any effect; otherwise, the charge on the gate capacitor $C_\mathrm{g}$ is conserved~\cite{ground}. Thus, the effective environment is $C_\mathrm{e}^{-1}=C_\mathrm{s}^{-1}+C_\mathrm{g}^{-1}$. A symmetric setup with a second inductance $L'$ between $C_\mathrm{s}$ and $C_2$ would create a third degree of freedom: a resonance with $\omega_\mathrm{res}^2=C_\mathrm{e}^{-1}[L^{-1}+(L')^{-1}]$, coupling to the Josephson phases $\phi_{1,2}$. Similarly if instead a $C_\mathrm{s}'$ would be inserted between $L$ and $C_1$ or parallel to $L$ or if, most physically, $C_\mathrm{s}$ were taken distributed along the ring. In all these alternatives, the additional high-frequency modes can be ignored at low energy and our conclusion would hold unchanged.

For the Hamiltonian one obtains ($\hbar=1$)
\beq\begin{split}
  H&=Q_1^2/2C_1+(Q_2{-}C_\mathrm{e}V_\mathrm{g})^2\!/2(C_2{+}C_\mathrm{e})-E_1\cos\phi_1 \\
    &\quad-E_2\cos\phi_2+(\phi_1{+}\phi_2{-}2\pi\Phi_\mathrm{x}/\Phi_0)^2\!/8e^2L\;;
\end{split}\eeql{H}
$\Phi_0$ is the flux quantum. $Q_2$, the conjugate to $\phi_2$, has a contribution on $C_\mathrm{s}$ besides the charge across $C_2$. Our basic environment thus merely renormalizes $C_2$~\cite{renorm}. Now put $\phi\equiv\phi_1{+}\phi_2$, $2\theta\equiv\phi_1{-}\phi_2$, implying $2Q_\phi=Q_1{+}Q_2$, $Q_\theta=Q_1{-}Q_2$. The $Q_{\phi,\theta}$ capacitance matrix is $\mathbb{C}=\Bigl(\begin{smallmatrix} \frac{1}{4}(\tilde{C}_1{+}\tilde{C}_2) & \frac{1}{2}(\tilde{C}_1{-}\tilde{C}_2) \\ \frac{1}{2}(\tilde{C}_1{-}\tilde{C}_2) & \tilde{C}_1{+}\tilde{C}_2 \end{smallmatrix}\Bigr)$, with $\tilde{C}_1=C_1$ and $\tilde{C}_2=C_2{+}C_\mathrm{e}$. At least in this model, the diagonal entries of $\mathbb{C}$ are still related. More importantly, $\mathbb{C}$ is not diagonal as the effective $\tilde{C}_1$ and $\tilde{C}_2$ differ in general---due to not only the environment, but mainly the inevitable fabrication spread. The cross-capacitance $C_{\phi\theta}$ adds a term $\propto C_{\phi\theta}\dot{\phi}\dot{\theta}$ to the Lagrangian~(4) in~\cite{FA}, so that the two tunneling paths in their Fig.~2 become inequivalent and no longer cancel. For an arbitrary ratio of Coulomb and Josephson energies, the symmetries proving level degeneracy~\cite{FA} require $E_1=E_2$ \emph{and} $\tilde{C}_1=\tilde{C}_2$. It would be interesting to see if and to what extent a $C_{\phi\theta}\neq0$ could be compensated for by a change in working point.

It may be useful to compare with the two-junction design in~\cite{saclay}, and clarify why $C_{\phi\theta}$ does not matter there but does in~\cite{FA}. In~\cite{saclay}, $\phi$ is made near-classical by a large junction and capacitor, so its conjugate charge and hence the latter's coupling to $Q_\theta$ can be ignored. Here, quantum effects in \emph{both} phases are essential for Aharonov--Casher type interference. Indeed, the device of~\cite{FA} is a modified rf SQUID, and suppressing overall-phase fluctuations in the latter would merely yield trivial classical dynamics.

In closing, let us point out that also note~22 in~\cite{FA} is incorrect. A superconducting loop with only two junctions is not bistable for all $L$, since for $L\rightarrow0$ the Josephson coupling is a simple cosine~\cite{bist-note}. This is precisely the reason why the design in~\cite{3JJ}, which does not rely on magnetic energy to achieve bistability, involves \emph{three} junctions.

I thank M.H.S. Amin, A. Blais, A. Shnirman, and A.M. Zagoskin for helpful comments.


\begin{thebibliography}{9}
\bibitem{ground} However, while one must have $C_\mathrm{s}\neq0$, grounding the loop, i.e.\ $C_\mathrm{s}\rightarrow\infty$, is admissible.
\bibitem{renorm} Clearly, in more general circuit models such as mentioned below Fig.~\ref{fig1} also $C_1$ gets renormalized. However, complete suppression of tunneling always requires fine-tuning two effective capacitances to be exactly equal.

\newpage
\bibitem{bist-note} This cosine's amplitude vanishes iff $E_1=E_2$ and $\Phi_\mathrm{x}=\Phi_0/2$, in which case higher-order terms determine the potential landscape and the minima in Fig.~2 of~\cite{FA} persist for small~$L$. However, the barrier separating these minima vanishes for $L\rightarrow0$, so that they are not ``metastable flux states'' in any operational sense.
\end{thebibliography}
\end{document}